\documentclass[twocolumn, showpacs, aps, pre, superscriptaddress, amsmath, amssymb, nofootinbib]{revtex4-1}

\usepackage{xcolor}
\usepackage{float}
\usepackage[utf8]{inputenc}
\usepackage{graphicx}
\usepackage{amsmath}

\usepackage[linktocpage,colorlinks=true,allcolors=blue,breaklinks=true]{hyperref}

\usepackage{ulem}
\usepackage[title]{appendix}
\usepackage{enumitem}
\usepackage{mathtools}
\usepackage[caption=false]{subfig}

\renewcommand{\vec}{\boldsymbol}
\newcommand{\diff}{\mathrm{d}}

\newcommand{\abs}[1]{|#1|}
\newcommand{\mean}[1]{\langle#1\rangle}
\newcommand{\Eqref}[1]{Eq. \eqref{#1}}
\newcommand{\figref}[1]{Fig. \ref{#1}}
\newcommand{\Eqsref}[1]{Eqs. \eqref{#1}}
\newcommand{\phiEq}{\phi^\mathrm{eq}}
\newcommand{\phiEqIn}{\phiEq_\mathrm{in}}
\newcommand{\phiEqOut}{\phiEq_\mathrm{out}}
\newcommand{\phiShell}{\phi_\mathrm{shell}}
\newcommand{\jIn}{\vec{j}_\mathrm{in}}
\newcommand{\jOut}{\vec{j}_\mathrm{out}}

\newcommand{\phiOut}{\phi_\mathrm{out}}
\newcommand{\vn}{v_\mathrm{n}}
\newcommand{\dx}{\Delta x}
\newcommand{\ds}{\Delta s}
\newcommand{\dt}{\Delta t}
\newcommand{\kOut}{k_\mathrm{out}}
\newcommand{\kf}{k_\mathrm{f}}
\newcommand{\kb}{k_\mathrm{b}}

\begin{document}

\title{Effective simulations of interacting active droplets}

\author{Ajinkya Kulkarni}
\affiliation{Max Planck Institute for Dynamics and Self-Organization, Am Fassberg 17, 37077 G{\"o}ttingen, Germany}

\author{Estefania Vidal-Henriquez}
\affiliation{Max Planck Institute for Dynamics and Self-Organization, Am Fassberg 17, 37077 G{\"o}ttingen, Germany}

\author{David Zwicker}
\email{david.zwicker@ds.mpg.de}
\affiliation{Max Planck Institute for Dynamics and Self-Organization, Am Fassberg 17, 37077 G{\"o}ttingen, Germany}

\date{\today}

\begin{abstract}
{
Droplets form a cornerstone of the spatiotemporal organization of biomolecules in cells. These droplets are controlled using physical processes like chemical reactions and imposed gradients, which are costly to simulate using traditional approaches, like solving the Cahn-Hilliard equation. To overcome this challenge, we here present an alternative, efficient method. The main idea is to focus on the relevant degrees of freedom, like droplet positions and sizes. We derive dynamical equations for these quantities using analytical solutions to simplified situations. We verify our method against fully-resolved simulations and show that it can describe interacting droplets under the influence of chemical reactions and external gradients using only a fraction of the computational costs of traditional methods. Our method can be extended to include other processes in the future and will thus serve as a relevant platform for understanding the dynamics of droplets in cells.
}

\end{abstract}


\maketitle

\section{Introduction}

Phase separation has recently been recognized as a powerful mechanism to organize biomolecules in the interior of biological cells~\cite{Keating2021,Mitrea2016,Banani2017,Hyman2014,Lyon2021}.
The droplets that spontaneously form via phase separation allow cells to sort molecules into compartments, which facilitates different functions, including controlling reactions~\cite{ANDERSSON2008275,Mingjian2018}, storing molecules~\cite{Greenblatt2018}, and buffering stochastic noise~\cite{Klosin2020}. %
To control these processes, cells need to regulate phase separation in space and time.
Examples for control mechanisms include chemical gradients \cite{Brangwynne2009, Saha2016,Weber2017,Wu2018}, chemical modifications of the involved molecules \cite{Hofweber2019,Ryan2018,QAMAR2018720,NOTT2015936,Zwicker2014, Soeding2019, Kirschbaum2021}, and global parameters, like pH and temperature \cite{Orij2011,Kroschwald2018,Peters2013,Petrovska2014,Fritsch2021,Putnam2019}.
Numerical simulations offer an attractive way to investigate these physical systems to understand how cells control their many droplets.

The dynamics of droplets are often simulated using the Cahn-Hilliard equation \cite{Review2019,CahnHilliardEq,Cahn1958}.
This fourth-order partial differential equation is typically expensive to simulate since it requires fine spatial discretization and small time steps.
Earlier approaches have improved the computational speed of numerical simulations of the Cahn-Hilliard equation using multi-grid methods \cite{Lee2021}, finite element modeling \cite{zhou2015, Chen2021}, incorporating mesh-less methods \cite{MOHAMMADI2019919}, and adaptive grids \cite{BANAS20082,Ceniceros2007}, but the fundamental drawbacks still persist.
Other approaches, such as Molecular Dynamics simulations \cite{Wingreen2021, Schuster2021} and Monte-Carlo methods \cite{Jacobs2021}, have also been used to simulate phase separation, but are also computationally expensive since they resolve details that are often not necessary for predicting the dynamics of droplets.

In this paper, we present a fast and efficient numerical method for simulating the dynamics of many interacting droplets.
\footnote{The source code of the project is freely available under the \href{https://doi.org/10.5281/zenodo.6656012}{DOI 10.5281/zenodo.6656012}.}
The method is based on analytical results from a thin-interface approximation of the continuous Cahn-Hilliard equation~\cite{Zwicker2015}.
In our effective model, we describe only the dynamics of the necessary degrees of freedom, which are the droplet positions and radii as well as some coarse information about the dilute phase.
The interaction of the droplets via the dilute phase is captured by discretizing their vicinity into thin annular sectors.
The dynamics of droplet growth and drift follows from material fluxes exchanged between the droplet and the dilute phase, which are obtained from solving a steady-state reaction-diffusion equation inside all sectors.
We present the model by first introducing the basic thermodynamic principles of phase separation, then the analytical theory behind the effective droplet model, and finally the details of the numerical method describing the dynamics of the droplets and dilute background.

\section{Model}

\label{sec:Model}

The main idea of our model is to replace the detailed description of the entire concentration field by the relevant degrees of freedom of the droplets.
We focus on the typical situation of well-separated droplets that are spherical due to surface tension, and thus describe the droplets by their positions $\vec{x}_i$ and radii $R_i$.
To build up the theory systematically, we will next introduce the continuous theory, the analytical description of isolated droplets, and then the full effective model.

\subsection{Continuous theory of phase separation}

We consider an isothermal, incompressible fluid in a closed system of volume $V_\mathrm{tot}$ that consists of solvent and droplet material.
The composition is described by the volume fraction $\phi(\vec{x}, t)$ of the droplet material, while the solvent volume fraction is given by $1-\phi$.
The thermodynamic state of the system is governed by the free energy
$ F[\phi] = \int \left ( f(\phi) + \frac{\kappa}{2} \, |\boldsymbol{\nabla} \phi| ^ 2 \right ) \mathrm{d}V$; see \cite{CahnHilliardEq},
where $\kappa$ is a parameter related to the interfacial tension, while phase separation is promoted by the free energy density $f(\phi)$.
For simplicity, we here consider a polynomial form,
\begin{equation}\label{eqn:free_energy}
    f(\phi)  = \frac{b}{2}(\phi - \phi^{0}_\mathrm{out})^2(\phi - \phi^{0}_\mathrm{in})^2 \;,
\end{equation}
where the minima $\phi^{0}_\mathrm{in}$ and $\phi^{0}_\mathrm{out}$ are the equilibrium concentrations in a thermodynamically large system and $b$ denotes the energy scale.
The dynamics of the system follow from the continuity equation, $\partial_t \phi + \boldsymbol{\nabla} \cdot \vec{j} = s$.
Here, $\vec{j}$ denotes spatial fluxes, which for simplicity will only be driven by diffusive processes, so hydrodynamic fluxes are neglected.
Conversely, $s$ is a source term related to chemical reactions that convert droplet material into solvent and vice versa \cite{Zwicker2015}.
Chemical reactions are typically local and often described by rate laws that depend on composition.
Conversely, the non-local diffusive fluxes $\vec j$ are driven by gradients in chemical potential
\begin{equation}
\label{eqn:mu}
    \mu = \nu \frac{\delta F}{\delta \phi} = b (\phi^{0}_\mathrm{in} - \phi) (\phi^{0}_\mathrm{out} - \phi) (2\phi - \phi^{0}_\mathrm{in} - \phi^{0}_\mathrm{out}) - \kappa \boldsymbol{\nabla}^2 \phi \;,
\end{equation}
where $\nu$ is the molecular volume of the droplet material.
Linear non-equilibrium thermodynamics implies $\vec{j} = -\Lambda(\phi) \boldsymbol{\nabla}\mu$, where $\Lambda(\phi)$ is a positive mobility \cite{GrootBook}.
Hence,
\begin{equation} \label{eqn:CHActive}
    \frac{\partial \phi}{\partial t} = \boldsymbol{\nabla} \cdot [\Lambda(\phi) \boldsymbol{\nabla} \mu] + s(\phi) \;,
\end{equation}
is a fourth-order, non-linear partial differential equation requiring two boundary conditions.
We here focus on the typical choice of no-flux conditions ($\vec{n} \cdot \boldsymbol{\nabla} \mu = 0$) and that solvent and droplet material interact identically with the system's boundaries ($\vec{n} \cdot \boldsymbol{\nabla} \phi = 0$), where $\vec{n}$ denotes the normal vector at the boundary.

In the case without chemical reactions ($s=0$), \Eqref{eqn:CHActive} reduces to the seminal Cahn-Hilliard equation \cite{CahnHilliardEq}, which describes passive phase separation.
In particular, two bulk phases with composition $\phi^{0}_\mathrm{in}$ and $\phi^{0}_\mathrm{out}$ typically emerge.
These phases are separated by an interface of width $w = 2\sqrt{\kappa / b}$ with surface tension $\gamma = \frac16 \sqrt{b/ \kappa}$ \cite{Zwicker2015,Review2019}.
When chemical reactions are weak, this general structure is typically preserved, although long-term dynamics, like Ostwald ripening, can be strongly modified \cite{Zwicker2015}.
Strong chemical reactions can actually destroy droplets \cite{Zwicker2015} and they might also lead to more complicated patterns \cite{Glotzer1995,Christensen1996}, which go beyond the scope of this paper.
Instead, we here focus on situations where well-defined droplets with a thin interface are typical.

The dynamics described by \Eqref{eqn:CHActive} adequately describes phase separation, but it can be prohibitively costly to simulate due to multiple reasons:
(i) The interface needs to be resolved, implying discretizations on the order of typically small interface width $w$.
(ii) The equation contains fourth-order derivatives in space, which often limits the time steps.
(iii) Interesting dynamics often take place on very long time scales.
For instance, during Ostwald ripening, length scales in the system evolve as $t^{ 1/3}$ \cite{Lifshitz,Wagner}, requiring long simulations to capture relevant behaviour.
However, since we are interested primarily in modelling dynamics of droplets, we circumvent these problems by focusing only on phase separation inside the nucleation and growth regime of the free energy density, and assume that sufficiently finite perturbations have already nucleated droplets.
We employ the thin-interface approximation \cite{Zwicker2015}, which is a coarse-grained analytical formulation of the continuous model (\Eqref{eqn:CHActive}) valid when the system is subject to strong phase separation, low variation of volume fractions in the droplet phase and the dilute phase, and large droplet sizes compared to the interface width.
This analytical approach was utilized earlier to study kinetics of many droplet systems and effects of chemical reactions on such systems \cite{Zwicker2015}.

In the next section, we elaborate on using the thin-interface approximation to build an effective droplet model describing the dynamics of droplets and dynamics of the dilute phase separately, instead of the full volume fraction field from the continuous model, thus effectively `de-coupling' the description of phase separated droplets from the dilute phase.

\subsection{Effective description of isolated droplets}

To build the effective model, we next derive approximate descriptions of the dynamics of the radius $R$ and position $\vec{x}$ of an isolated droplet.
Since we only consider spherical droplets with a thin interface ($R \gg w$) and weak chemical reactions, we can use basic thermodynamics to derive the equilibrium concentrations inside and outside of the interface of the droplet, denoted by $\phiEq_\mathrm{in}$ and $\phiEq_\mathrm{out}$, respectively.
Due to surface tension effects, they are slightly elevated above the basal values $\phi^0_\mathrm{in}$ and $\phi^0_\mathrm{out}$ prescribed by the free energy density; see \Eqref{eqn:free_energy},
To first order in the curvature of the surface, we have
\begin{subequations}
\label{eqn:equilibrium_conditions}
\begin{align}
    \phiEq_\mathrm{in} &= \phi^{0}_\mathrm{in} \left (1 + \frac{l_{\gamma, \mathrm{in}}}{R} \right )
    \qquad \text{and}
\\[10pt]
    \phiEq_\mathrm{out} &= \phi^{0}_\mathrm{out} \left ( 1 + \frac{l_{\gamma, \mathrm{out}}}{R}\right )
    \;,
\end{align}
\end{subequations}
where $l_{\gamma, \mathrm{out}}$ and $l_{\gamma, \mathrm{in}}$ are capillary lengths \cite{Review2019}.
In our case they read $l_{\gamma, \mathrm{in}} = (\kappa/b)^{1/2} /[3 \phi^{0}_\mathrm{in} \left(\phi^{0}_\mathrm{in} - \phi^{0}_\mathrm{out}\right)^{3}]$ and $l_{\gamma, \mathrm{out}} = (\kappa/b)^{1/2}/[3 \phi^{0}_\mathrm{out} \left(\phi^{0}_\mathrm{in} - \phi^{0}_\mathrm{out}\right)^{3}]$ in three dimensions.
The dynamics of the volume fraction field $\phi_\mathrm{in}$ inside the droplet is in principle described by \Eqref{eqn:CHActive}, but since the composition typically hardly varies, we can linearize $\phi_\mathrm{in}$ around $\phi^{0}_\mathrm{in}$ to obtain

\begin{align}
    \label{eqn:RD_droplet}
    \frac{\partial \phi_\mathrm{in}}{\partial t}
        \approx D_\mathrm{in} \boldsymbol{\nabla} ^2 \phi_\mathrm{in} +
        s(\phi^{0}_\mathrm{in})
        - k_{\mathrm{in}}(\phi_\mathrm{in} - \phi^{0}_\mathrm{in}) \;,
\end{align}
where $D_\mathrm{in} = \Lambda(\phi^0_\mathrm{in}) \, b$ is the diffusivity and $k_{\mathrm{in}}= - s'(\phi^{0}_\mathrm{in})$ denotes the reaction rate \cite{Review2019}.
Generally, positive rates ($k_\mathrm{in}>0$) stabilize the volume fraction $\phi_\mathrm{in}$, while negative rates might destabilize it.
However, the instability is suppressed when the droplet radius $R$ is small compared to the reaction-diffusion length scale, $\xi_\mathrm{in}=\sqrt{D_\mathrm{in}/|k_{\mathrm{in}}|}$ \cite{Review2019}.
Since we here consider weak chemical reactions, $\xi_\mathrm{in}$ will be large, and we thus assume $R \ll \xi_\mathrm{in}$ in the following.
We use this to solve \Eqref{eqn:RD_droplet} in stationary state in a system with angular symmetry using the boundary conditions $\phi_\mathrm{in}(R) = \phiEq_\mathrm{in}$ and $\partial_r \phi_\mathrm{in}(0) = 0$.
The analytical result allows us to estimate the diffusive flux $\vec{j}_\mathrm{in}$ inside the interface,
\begin{equation}
	\label{eqn:flux_inside}
    \vec{j}_\mathrm{in} \approx \frac{R}{d}  s(\phi^\mathrm{eq}_\mathrm{in})\,\vec{n}
    \;,
\end{equation}
where $d$ is the space dimension; see Appendix \ref{sec:fluxes_inside_droplet}.
Production of droplet material inside the droplet ($s(\phi^\mathrm{eq}_\mathrm{in}) > 0$) leads to an outward flux $\vec{j}_\mathrm{in} \cdot \vec{n} > 0$, which can drive droplet growth.
Conversely, destroying droplet material ($s(\phi^\mathrm{eq}_\mathrm{in}) < 0$) promotes shrinking droplets.
Droplets might also grow if they take up material from the surrounding. 
Similarly to inside of droplets, the volume fraction $\phiOut$ will typically vary only little, so we can linearize around the base value $\phi^0_\mathrm{out}$ to obtain the reaction-diffusion equation
\begin{equation}
	\label{eqn:RD_dilute}
	\frac{\partial \phiOut}{\partial t} \approx
	    D_\mathrm{out} \boldsymbol{\nabla} ^2 \phiOut + s(\phiOut)
	\;,
\end{equation}
where $D_\mathrm{out} = \Lambda(\phi^0_\mathrm{out}) \, b$ is the diffusivity outside droplets.
Solving this equation and obtaining the corresponding flux $\jOut$ outside the droplet is more difficult since the environment of the droplet might not be isotropic.
We thus discuss the coupling of droplets to the dilute phase $\phiOut$ in more detail in the next section.

If we know the fluxes $\jIn$ and $\jOut$, we can determine the net accumulation of droplet material at the interface, which implies droplet growth.
Note that only the normal components of the fluxes affect the shape, while the tangential components merely distribute material parallel to the interface.
The shape changes of an isolated droplet are thus described by the interfacial speed $\vn$ in the normal direction \cite{Review2019},
\begin{equation}
	\label{eqn:interfacefluxes}
	\vn \approx \frac{\jIn - \jOut}{\phiEqIn - \phiEqOut} \cdot \vec{n}
	\;;
\end{equation}
see Appendix \ref{sec:interfacial_velocity}.
General shape changes can result in non-spherical droplets, but since surface tension effects typically ensure a near-spherical shape, we project the general shape onto the degrees of freedom that we use to describe the droplet,
\begin{subequations}
\label{eqn:droplet_dynamics}
\begin{align}
	\label{eqn:DropletGrowth}
	\frac{\mathrm{d}R}{\mathrm{d}t} & = \frac{1}{S} \int \vn \, \diff A
	\qquad \text{and}
\\[10pt]
	\label{eqn:DropletDrift}
	\frac{\mathrm{d} \vec{x}}{\mathrm{d}t} &= \frac{d}{S} \int \vn \, \vec{n} \, \diff A
	\;,
\end{align}
\end{subequations}
where the integral is over the droplet surface, $d$ is the space dimension, and $S$ is the surface area of the droplet; see Appendix \ref{sec:droplet_dynamics}.
To summarize, \Eqsref{eqn:droplet_dynamics} determine how an isolated droplet evolves in time.
This involves \Eqsref{eqn:equilibrium_conditions}, \Eqref{eqn:flux_inside}, \Eqref{eqn:interfacefluxes} as well as an approximation for the fluxes $\jOut$ outside the droplet interface, which is the central part of our method that we discuss next.

\subsection{Numerical model for many droplets}

\begin{figure}[tb]
\centering
{\includegraphics[scale=0.24]{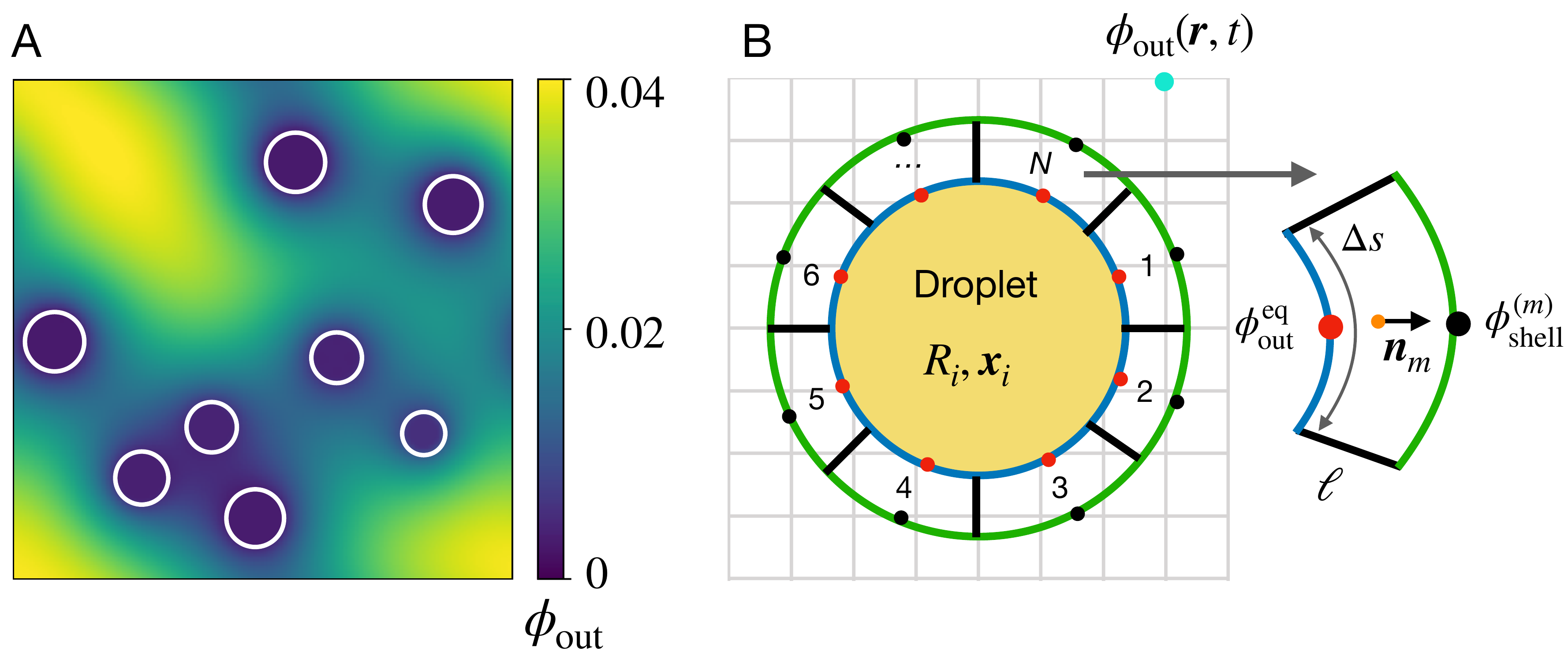}\label{fig:simulation}}
\caption{\textbf{Schematics of the simulation model, describing droplets and the background field $\phiOut$.}
(A) Droplets co-exist with the background field $\phiOut$ and interact with the it only through material fluxes.
For simplicity, $\phiOut$ also exists at the location of the droplets, but this has negligible effect on the dynamics.
(B) Isolated droplet (yellow) with a surrounding shell of thickness $\ell$, which is further discretized in $N$ sectors of linear size $\ds$.
The exchanged fluxes between droplet and background are determined in each sector based on the equilibrium fraction $\phiEqOut$ (red dots) and the value $\phiOut^{(m)}$ at the outer side (black dot), which is determined from the background field using bilinear interpolation.
The background field $\phiOut$ is uniformly discretized on a Cartesian grid (gray grid).
}
\label{fig:schematics}
\end{figure}

The dynamics of many droplets in the same system are coupled since they may exchange material via the dilute phase.
To describe this exchange, and ultimately derive the flux $\jOut$ at each droplet, we first consider the dynamics of the volume fraction of the dilute phase $\phiOut$.
In principle, the dynamics of $\phiOut$ follows from \Eqref{eqn:RD_dilute}, with appropriate boundary conditions applied at the system's boundary and at all droplet surfaces.
To simplify the description of the dilute phase, we assume that $\phiOut$ is defined in the entire system, including where droplets are; see \figref{fig:schematics}A.
In this picture, droplets are local perturbations that exchange material with the background field $\phiOut$.

\captionsetup[subfigure]{labelformat=empty}

\subsubsection{Dynamics of the background field}

The background field $\phiOut$ changes due to diffusion and reactions, even if droplets are absent.
To capture this dynamics, we discretize the continuous field $\phiOut$ on a uniform Cartesian grid with a distance $\Delta x$ between the neighbouring support points.
We then evolve \Eqref{eqn:RD_dilute} in time using finite central differences and explicit temporal stepping \cite{Zwicker2020}.
Since we do not need to resolve droplets at this scale, the spatial discretization can be much larger than in traditional Cahn-Hilliard equations.

\subsubsection{Growth of a single droplet in a background field}

To obtain the flux $\jOut$ in the vicinity of an isolated droplet, we need to determine $\phi_\mathrm{out}$ in the region surrounding this droplet.
We do this by considering \Eqref{eqn:RD_dilute} in an annular shell of thickness $\ell$ surrounding the droplet, which we further discretize into $N$ sectors in the angular dimensions; see Fig. \ref{fig:schematics}.
We choose the distribution of these angular sectors as uniformly as possible, correcting for potential asymmetries as described below.
For simplicity, we assume that fluxes in the angular directions are negligible, so we can express the volume fraction in the $m$-th sector as $\phiOut^{(m)}(r)$, where $r$ is a radial coordinate measuring distance from the droplet position $\vec{x}$.
We determine $\phiOut^{(m)}(r)$ using \Eqref{eqn:RD_dilute} in stationary state with the boundary conditions $\phiOut^{(m)}(R)=\phiEqOut$ and $\phiOut^{(m)}(R + \ell)= \phiShell^{(m)}$.
Here, $\phiShell^{(m)}$ is the volume fraction of droplet material in the background field at the outer side of the $m$-th shell sector, which we estimate from a linear interpolation of the discretized background field $\phi_\mathrm{out}$; see Fig. \ref{fig:schematics}.
Since $\phiOut^{(m)}(r)$ typically varies only marginally in the shell sector, we also linearize the reaction flux, $s(\phiOut^{(m)}) \approx \Gamma_\mathrm{out} - \kOut \,  \phiOut$, by imposing $s(\phiOut^{(m)})(R) = s(\phiEqOut)$ and $s(\phiOut^{(m)})(R + \ell) = s(\phiShell^{(m)})$.
This implies $\Gamma_\mathrm{out} = [\phiShell^{(m)} \, s(\phiEqOut) - \phiEqOut \, s(\phiShell^{(m)})] / (\phiShell^{(m)} - \phiEqOut)$ and $\kOut = [s(\phiEqOut) - s(\phiShell^{(m)})] / (\phiShell^{(m)} - \phiEqOut)$.
Taken together, we obtain an analytical approximation of $\phiOut^{(m)}(r)$ in each shell sector, from which we determine the local normal flux $\jOut^{(m)}$ outside the droplet; see Appendix \ref{sec:flux_expressions}.
Using these expressions together with \Eqsref{eqn:flux_inside}, \eqref{eqn:interfacefluxes} and \eqref{eqn:droplet_dynamics}, we find that individual droplets evolve according to
\begin{subequations}
\label{eqn:DropletDiscretized}
\begin{align}
    \frac{\mathrm{d}R}{\mathrm{d}t} &\approx \frac{1}{\phiEqIn} \sum_{m=1}^{N} \frac{A_m}{S}\left ( \frac{R}{d} s(\phi^\mathrm{eq}_\mathrm{in})  -  {j}^{(m)}_\mathrm{out}  \right)
    \qquad \text{and}
\\[10pt]
   \frac{\mathrm{d}\vec{x}}{\mathrm{d}t} &\approx \frac{d}{\phiEqIn} \sum_{m=1}^{N} \frac{A_m}{S}
    	\left ( \frac{R}{d}s(\phi^\mathrm{eq}_\mathrm{in})  -  {j}^{(m)}_\mathrm{out}\right) \vec{n}_m \;,
\end{align}
\end{subequations}
where $R$, $S$ are radius and surface area of the droplet, respectively, $A_m$ is the inner area of the shell sector and $\vec{n}_m$ is the unit vector pointing from the droplet center to the $m$-th shell center; see \figref{fig:schematics}.
We use \Eqsref{eqn:DropletDiscretized} to describe how internal reactions and external material exchange with the background affects the dynamics of each droplet.

\subsubsection{Coupled dynamics of droplets and the background field}

\Eqsref{eqn:DropletDiscretized} describe how droplets change when they exchange droplet material with the background field $\phiOut$.
Due to material conservation, the material flux from the droplet toward each sector~$m$, $\jOut^{(m)} \cdot \vec{n} \, A_m$, needs to accumulate in the background field.
We use a linear interpolation at the midpoint of the inner boundary of the shell section (red points in \figref{fig:schematics}B) to add the respective amount to the background field $\phiOut$.
Note that negative fluxes $\jOut^{(m)}$ distribute material from the background field to the droplet and thus lead to growth.
Taken together, this procedure ensures material conservation while preserving anisotropies of the exchange.

\subsubsection{Full simulation}
The full numerical method evolves the state of the system, i.e., the discretized background field $\phiOut(\vec{r})$ and the positions $\vec{x}_i$ and radii $R_i$ of all droplets, in time.
We propose an explicit iteration, where the state at $t+\Delta t$ is directly determined from the state at time $t$.
Here, we first evolve the reaction-diffusion equation \Eqref{eqn:RD_dilute} of the background field and then iterate over all droplets.
For each droplet, we determine the fluxes $\jOut^{(m)} \cdot \vec{n}$ for all shell sectors $m$ and remove the associated material from the background field.
We then update the droplet's position and radius according to \Eqref{eqn:DropletDiscretized}.
Starting from an initial state at $t=0$, this algorithm allows us to evolve the dynamics forward in time.

\subsection{Choosing simulation parameters}

The algorithm described above has several parameters that need to be chosen wisely for an accurate and fast simulation.
In particular, we need to specify the discretization $\dx$ of the background field, the shell thickness $\ell$, the typical size $\ds$ of a shell sector, and the time step $\dt$.
We next discuss suitable values for all four parameters using detailed simulation of the continuous model given by \Eqref{eqn:CHActive} as ground truth.
We later show that the resulting choice for the parameters can recapitulate many effects that have been described previously in the literature.

To calibrate the simulation parameters, we compare our effective model to the established continuous model given by \Eqref{eqn:CHActive}.
Our test case consists of two passive droplets of initial radius $R_0=20 w$ whose centers are separated by $S_\mathrm{d}=10 R_0$.
The droplets are placed in a background of vanishing initial volume fraction, $\phiOut(\vec{r}, 0) = 0$, so the system is under-saturated and the droplets will shrink.
For the continuous model, we exploit the angular symmetry of the problem and consider an azimuthally symmetric cylindrical domain where $r,z \in [0, 682 w]$.
Conversely, our effective model is simulated in a $3$-dimensional cubic domain of size $[0, 1000 w]^{3}$.
We compare the final radii of the droplets after a duration $T$, where the droplets typically have shrunk by about $20\%$.
The deviation of the mean droplet radius $\mean{R_*}$ of our effective model compared to the radius $\mean{R_\mathrm{CM}}$ of the continuous model allows us to determine the crucial simulation parameters $\dx$, $\ell$, and $\ds$.

\subsubsection{Grid discretization $\dx$}

The spatial discretization $\dx$ determines the resolution at which variations of the background field $\phiOut$ are resolved.
Consequently, the choice of $\dx$ is based on the problem:
If spatial interactions are negligible and a mean-field model is desired, $\dx$ can be arbitrarily large.
Conversely, if spatial correlations between droplets are important, $\dx$ needs to be smaller than the droplet separation.
Another case are external gradients that affect droplets \cite{Weber2017}, where $\dx$ needs to be on the order of the droplet radii, so spatial anisotropies can be resolved on the droplet level.

\subsubsection{Annular shell thickness $\ell$}

\begin{figure}[tb]
\centering
\includegraphics[scale=0.235]{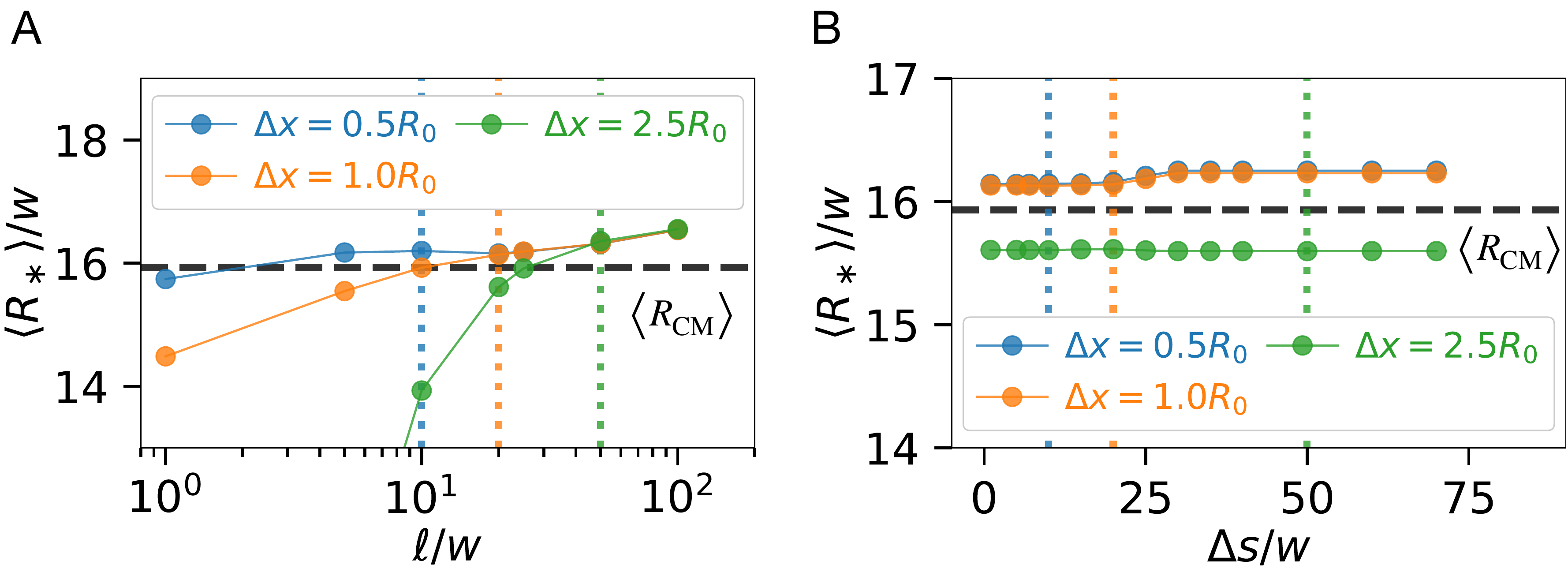}
\caption{
\textbf{Effect of annular shell thickness $\ell$ and sector size $\ds$ in simulations of a passive droplet pair.}
(A)~Mean droplet size $\langle R_\ast \rangle$ (filled dots) as a function of $\ell$ for various $\dx$ (vertical dotted lines) using $\ds \approx R_0$.
The ground truth $\mean{R_\mathrm{CM}}$ (black dashed line) is obtained from the continuous model.
(B)~$\langle R_\ast \rangle$ as a function of $\ds$ (filled dots) for various $\dx$ using $\ell \approx R_0$, with $\mean{R_\mathrm{CM}}$ shown as the black dashed line.
\mbox{(A, B)}
Two droplets with radii $R_0 = 20 w$ are placed with their centers $10 R_0$ apart in an empty background, $\phiOut(\vec{r}, t=0) = 0$.
Continuous simulations given by \Eqref{eqn:CHActive} used an azimuthally symmetric cylindrical domain with bounds $r,z \in [0, 682 w]$ with a spatial discretization of $0.5w$.
Effective simulations used a $3$-dimensional cubic domain of size $[0, 1000 w]^3$ and $l_{\gamma, \mathrm{in}} = 0.166 w$.
Additional  parameters are $s=0$, $\Lambda = w^2 / b \tau$, $\phi^{(0)}_\mathrm{out} = 0$, $\phi^{(0)}_\mathrm{in} = 1$, $\tau = w^2/D_\mathrm{out}$, and $w = 2 \sqrt{\kappa / b}$.
}
\label{fig:shell_parameters}
\end{figure}

The most crucial part of our numerical method describes how material exchanges between the droplets and the background field.
To describe this exchange faithfully, we interpolate the background field in an annular shell around the droplet.
The thickness $\ell$ of this shell can thus be interpreted as an interpolation length scale and its value affects the accuracy of the simulation:
If $\ell$ is too small, the fluxes $\jOut^{(m)}$ are overestimated since they scale with $\ell^{-1}$; see Appendix \ref{sec:flux_expressions}.
Conversely, if $\ell \gg \dx$, the background field would not be evaluated in the vicinity of the droplet, so interactions cannot be captured correctly.
Taken together, we conclude that $\ell \sim \dx$ is a reasonable choice for the shell thickness.
Indeed, \figref{fig:shell_parameters}A shows that this choice leads to a faithful estimate of the droplet growth for various values of $\dx$.

\captionsetup[subfigure]{labelformat=empty}

\subsubsection{Shell sector width $\ds$}

To resolve spatial anisotropies around a droplet, we discretize the shell into $N$ sectors; see \figref{fig:schematics}B.
To obtain $\phiShell$ for each sector, the background field $\phiOut$ is interpolated once per sector. 
Consequently, a larger number of sectors leads to a finer discretization and potentially a higher accuracy at the expense of larger computational cost.
However, the accuracy is limited by the background discretization $\dx$ as increasing $N$ will have hardly any benefit if the distance $\Delta s$ between interpolation points is already smaller than $\dx$.
For $d=2,3$ dimensions, we respectively use $\ds \approx 2\pi R/N$ and $\ds \approx\sqrt{4 \pi R^2 / N}$; see \figref{fig:schematics}B.
Using $\ds \sim \dx$, we can solve these equations for $N$, so that we can determine the number of sectors for each droplet based on its instantaneous radius.
Note that this implies that the dynamics of larger droplets will be described by more sectors to faithfully capture the interaction with their surrounding.
\figref{fig:shell_parameters}B shows that the shell sector size $\ds$ has only marginal effects in simple situations.

\subsubsection{Time step $\dt$}

While the three previously discussed parameters affect material fluxes between droplets and background, the time step $\dt$ determines the accuracy of dynamics of the model.
Smaller values of $\dt$ imply more accurate simulations, while larger values result in faster simulations, although numerical instabilities might also render simulations unstable.
We next separately analyze the dynamics of the background field, the shell, and the droplet growth to identify the maximal suitable value of $\dt$.

The dynamics of the background are described by the partial differential equation \Eqref{eqn:RD_dilute}, which we here solve using a simple explicit Euler scheme.
A standard von Neumann stability analysis shows that this scheme is stable if $\dt < \dx^2/(2 D_\mathrm{out})$, where $D_\mathrm{out}$ is the diffusivity in the background field.
Consequently, a suitable time step for evolving the background field is $\dt_\mathrm{background} = 0.1 \dx^2/D_\mathrm{out}$, where we chose the constant pre-factor conservatively.
Similarly, we define $\dt_\mathrm{shell} = 0.1 \ell^2/D_\mathrm{out}$ for the shell.
To ensure faithful dynamics of droplet growth, we demand that the relative growth $R^{-1}\abs{\diff R/\diff t}$ is small during a single time step $\dt$.
Assuming that typical droplets are not much smaller than the mean initial droplet radius $\mean{R}$, this implies a maximal time step $\dt_\mathrm{drop} = 0.1 \mean{R}^2/D_\mathrm{out}$.
Finally, we also consider the time scale of reactions, $\dt_\mathrm{reaction} = 0.1 / (\max_\phi |s(\phi)|)$, based on the maximal rate of $s(\phiOut)$.
Taken together, we set the time step of the simulation to the minimal value of the four limiting time scales determined above.

\section{Validation}

We showed above that $\Delta x \approx \ell \approx \Delta s$ is a sensible choice for the parameters of our algorithm.
To see how this choice affects accuracy and speed of the simulation, we next present three simulation scenarios, which range from single droplets in an heterogeneous environment to droplet coarsening of large dilute emulsions.

\subsection{Passive droplet in external gradient}

\begin{figure}[tb]
\centering
\includegraphics[scale=0.247]{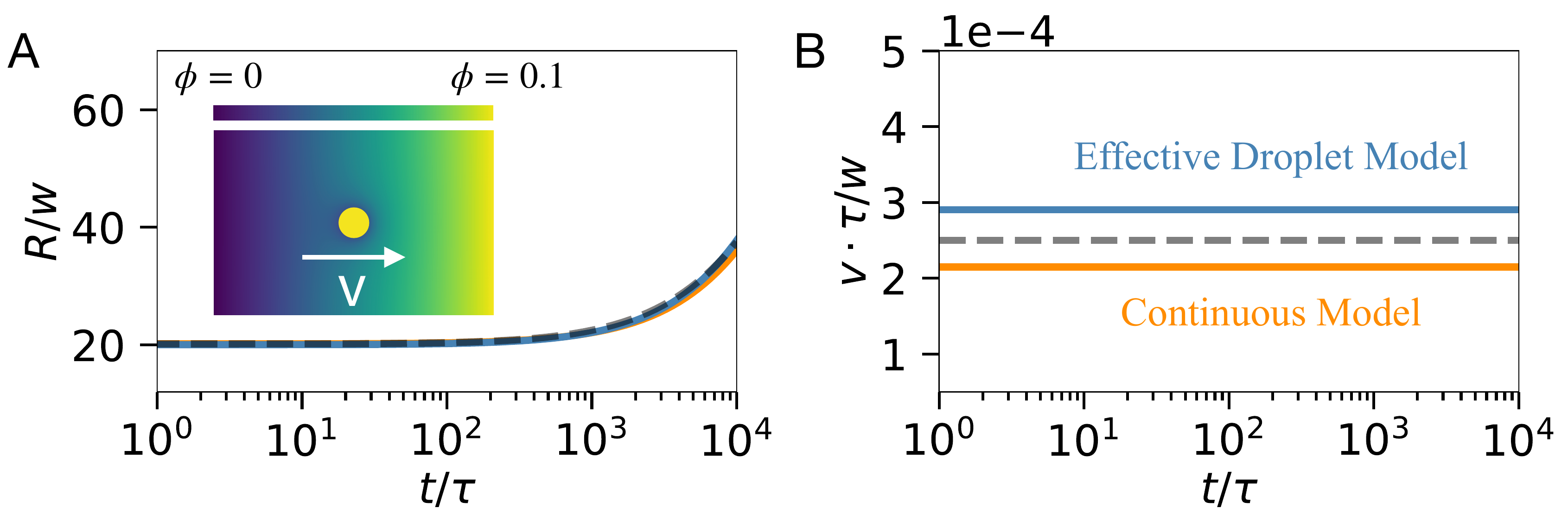}
\caption{\textbf{Droplet dynamics in external gradients.}
(A)\,Droplet radius $R$ as a function of time $t$ compared to the analytical prediction (dashed line) from the thin-interface approximation \cite{Review2019}, the effective droplet model (blue) and the continuous model (orange).
The inset shows a schematic of the simulation with the gradient imposed in the background.
(B) Droplet drift speed $v$ as function of $t$ compared to the analytical prediction (dashed line) from the thin-interface approximation \cite{Review2019}.
\mbox{(A, B)}
The continuous model uses a cylindrical domain with $r \in [0, 400 w]$, $z\in [-L_2, L_2]$ with $L_2=600 w$, azimuthal symmetry, and boundary conditions $\mu(z=-L_2) = 0$ and $\mu(z=L_2)= 0.072\,b w^3$ to impose the gradient.
The effective model uses a $3$-dimensional box of size $[-L_3, L_3]^3$ with $L_3 = 422 w$, $\dx = R_0$, $\ell \approx \Delta s \approx R_0$, and boundary conditions $\phiOut(y = -L_3) = 0.01483$ and $\phiOut(y = L_3) = 0.0851$.
Note that in the absence of droplets, boundary conditions imply identical linear gradient in the continuous model as $\phi = \phi(z)$ and in the effective droplet model as $\phiOut = \phiOut(y)$, which were also used to initialize the background for both models.
Remaining parameters are specified in Fig. \ref{fig:shell_parameters}.
}
\label{fig:drop_in_gradient}
\end{figure}

We first consider a single droplet in an external composition gradient, which is maintained via boundary conditions.
Biological cells use such a setup to control the position of droplets in their interior \cite{Brangwynne2009,Griffin2011,Weber2017}.
\figref{fig:drop_in_gradient} shows that the effective droplet model captures the drift and growth of the passive droplet quantitatively.
While the resulting dynamics are very similar, the run time of the simulations are very different: The continuous model took roughly one day to complete, while the effective model finished within 10 seconds on identical hardware.
Since the continuous model is much slower, we performed the following tests only with the effective model.

\captionsetup[subfigure]{labelformat=empty}

\subsection{Mean-field coarsening of passive droplets}

\begin{figure}[tb]
\centering
\includegraphics[scale=0.247]{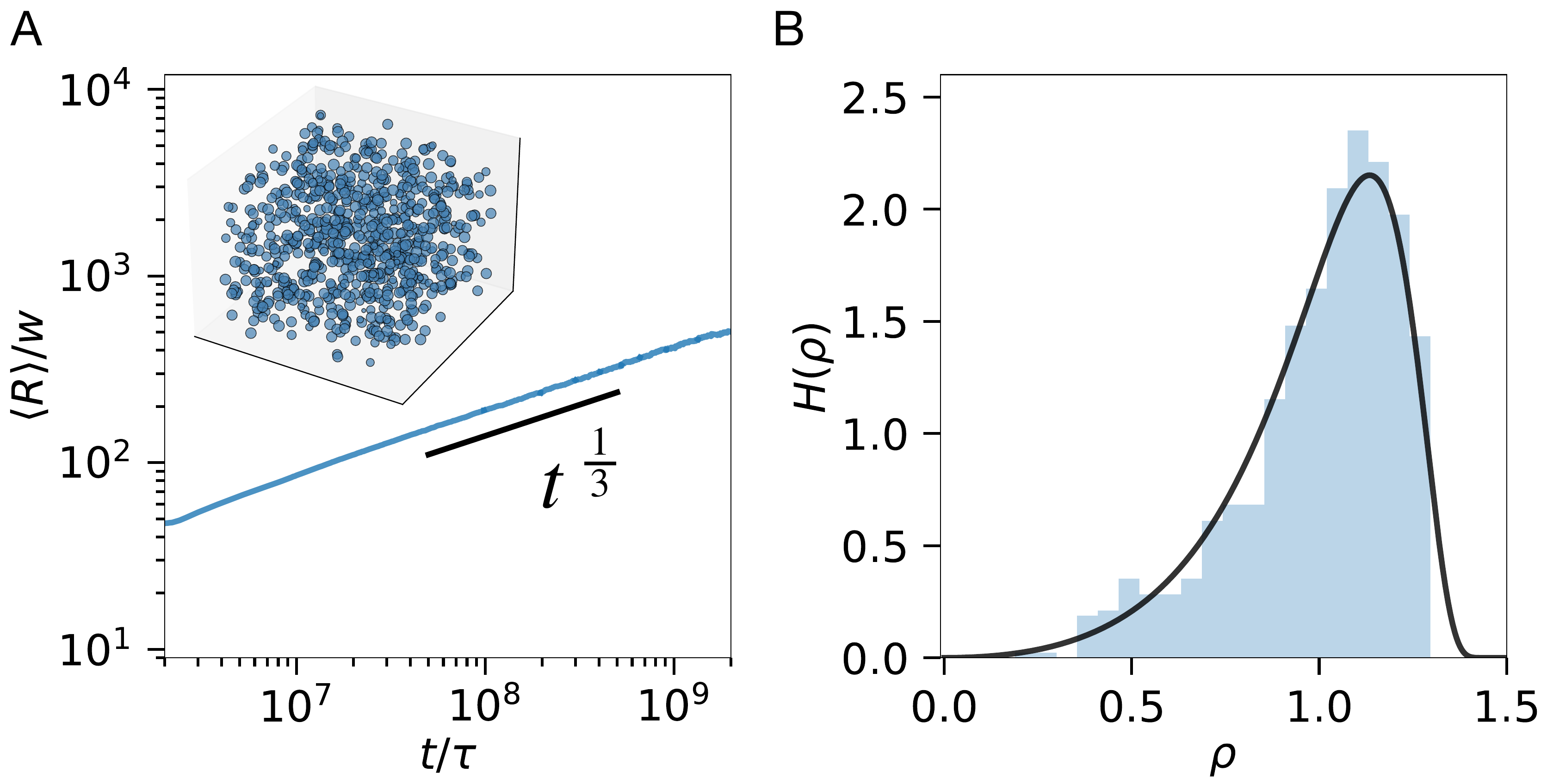}
\caption{\textbf{Ostwald ripening of passive droplets}.
(A) Mean droplet radius $\langle R \rangle$ as a function of time $t$ shows the expected scaling $\langle R \rangle \propto t^{1/3}$ \cite{Lifshitz,Wagner}.
Inset shows snapshot at $t=2 \times 10^8 \tau$.
(B) Frequency $H(\rho)$ of the normalized radius $\rho = R/\left \langle R \right \rangle$ at $t = 2 \times 10^8 \tau$ compared to the expected universal distribution (black)  \cite{Lifshitz,Wagner}.
(A, B)
Simulations were carried out in a $3$-dimensional periodic cubic domain of size $[0, L]^3$, where $L = 10^4 w$.
We used $\Delta x \approx \ell \approx L$ and a single shell sector to approach the mean field solution.
$10^5$ droplets were initialized with radii chosen uniformly in $[9.5 w, 10.5 w]$ in an initial background $\phiOut(t=0) = 0.05$.
Remaining parameters are specified in Fig. \ref{fig:shell_parameters}.}
\label{fig:passive_emulsions}
\end{figure}

We next consider the interactions of many passive droplets in a dilute emulsion.
When droplets only interact via the spatially averaged background field, Lifshitz and Slyozov predicted that the average droplet radius $\left \langle R \right \rangle$ grows as $t^{1/3}$ in this case \cite{Lifshitz,Wagner,LSWanalytics}.
Our simulation of $10^5$ droplets indeed recovers this scaling (\figref{fig:passive_emulsions}A) when we mimic this situation by setting the discretization $\dx$ to the system size.
Moreover, \figref{fig:passive_emulsions}B shows that the distribution of radii also follows the universal shape $H(\rho) = \frac{4}{9}\rho^2\left ( 1+ \frac{\rho}{3} \right ) ^ {-7/3} \left ( 1-\frac{2 \rho}{3} \right )^{-11/3} \mathrm{exp} \left ( 1 - \frac{3}{3 - 2\rho} \right )$, where $\rho=R/{\left \langle R \right \rangle}$ is a scaled droplet size \cite{Lifshitz}.
Our effective model thus faithfully captures the dynamics of many droplets, optionally even beyond the Lifshitz-Slyozov regime by increasing the spatial resolution to capture correlations in droplet growth.

\captionsetup[subfigure]{labelformat=empty}

\subsection{Mean-field coarsening of active droplets}

\begin{figure}[tb]
\centering
{\includegraphics[scale=0.247]{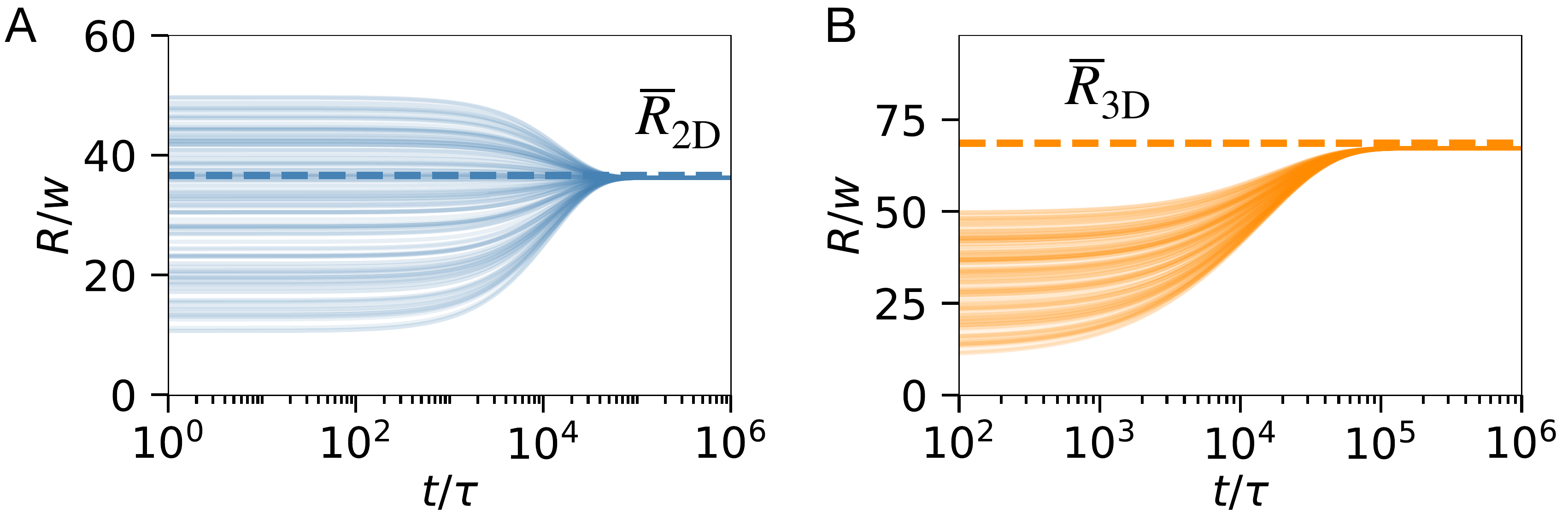}
\label{fig:active_emulsions}}
\caption{\textbf{Suppression of Ostwald ripening by first-order chemical reactions}.
(A, B) Radii $R$ as a function of time $t$ of droplets evolving in $d=2$ dimensions (A) and $d=3$ dimensions (B).
The theoretically expected radii are indicated by dashed horizontal lines.
$100$ droplets with radii chosen uniformly in $[10w, 50w]$ were placed in a periodic cubic domain of size $[0, L]^d$ with $L=1000 w$.
Model parameters are $s(\phi)= \kf (1 - \phi) - \kb \phi$, $\phiOut(t=0) = \kf/(\kf + \kb)$, $\kf = 10^{-5} \tau^{-1}$, $\kb = 10^{-4}  \tau^{-1}$, $\Delta x \approx \ell \approx L$, and a single shell sector for all droplets.
Remaining parameters are specified in Fig. \ref{fig:shell_parameters}.
}
\label{active_emulsions}
\end{figure}

As a final example, we consider the interaction of many active droplets in a dilute emulsion.
Here, we focus on the simplest case of a first-order reaction between the solvent and the droplet material, which is known to suppress Ostwald ripening \cite{Review2019,Zwicker2015}.
We thus solve \Eqref{eqn:CHActive} using the reaction flux $s(\phi) = \kf (1 - \phi) - \kb (\phi)$.
\figref{active_emulsions} shows that the emulsions with broad initial sizes quickly converge to mono-disperse distributions in $2$ and $3$ dimensions.
The droplet size in stationary state is very close to the theoretical prediction, which we obtain numerically from the condition $\jIn = \jOut$ using \Eqsref{eqn:RD_droplet} and \eqref{eqn:RD_dilute}.
Taken together, we thus demonstrated that our effective model faithfully recovers important physical behaviour of active droplets.

\captionsetup[subfigure]{labelformat=empty}

\section{Outlook and discussion}
\begin{figure}[tb]
\centering
\includegraphics[scale=0.238]{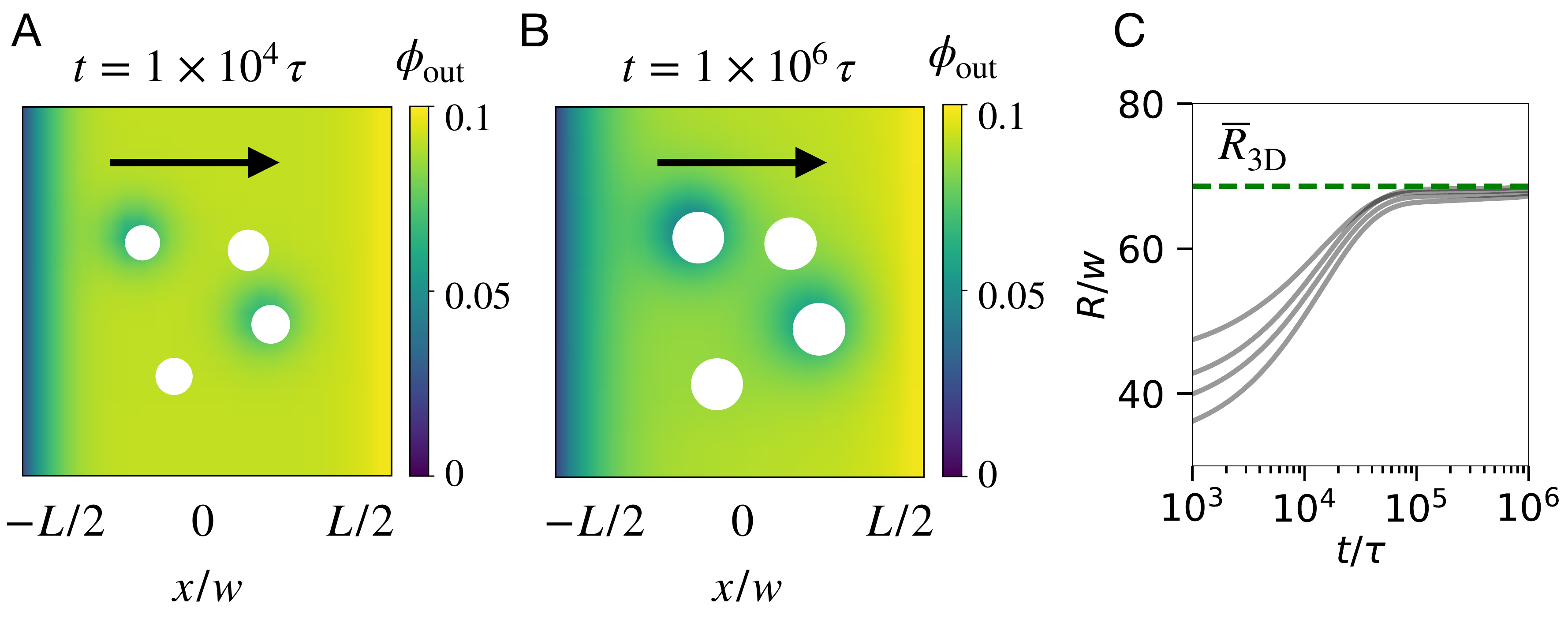}
\caption{\textbf{Active droplets in an external gradient.}
\mbox{(A-B)} Snapshots of two-dimensional projections of the three-dimensional system at times indicated above the frame, showing droplet drift up the gradient (black arrow).
White circles denote droplets while color indicates the volume fraction in the background field.
(C) Droplet radius $R$ as a function of time $t$ indicating that all droplets reach $\overline{R}_\mathrm{3D}$ (dashed line); compare to \figref{active_emulsions}.
(A--C) Four droplets with radii chosen uniformly from $[0.8 R_0, 1.2 R_0]$ were placed in a cubic Cartesian domain of size $[-\frac L2, \frac L2]^3$ with boundary conditions $\phiOut(x = -\frac L2) = 0$ and $\phiOut(x = \frac L2) = 0.1$ to impose the gradient and no-flux boundary conditions at the remaining system boundaries.
Model parameters are $L = 10^3 w$, $\dx \approx \ell \approx \Delta s \approx R_0 \approx 40 w$, $s(\phi)= \kf (1 - \phi) - \kb \phi$, $\phiOut(t=0) = \kf/(\kf + \kb)$, $\kf = 10^{-5} \tau^{-1}$ and $\kb = 10^{-4} \tau^{-1}$.
Remaining parameters are specified in Fig. \ref{fig:shell_parameters}.
}
\label{fig:active_emulsion_gradient}
\end{figure}

We showed that our effective method is orders of magnitudes faster than the continuous model while still accurately capturing the dynamics of droplets under the influence of chemical reactions and external gradients.
To demonstrate that the method also extends to more challenging situations, we finally simulate the combination of chemical reactions and external gradients on the dynamics of the droplets.
\figref{fig:active_emulsion_gradient} shows that droplets grow as they drift along the gradient and they approach the fixed radius given by $\overline{R}_\mathrm{3D}$, so that this system controls droplet drift and size.
Taken together, our simulations demonstrate that the novel simulation method captures the dynamics of interacting active droplets efficiently.

To gain the significant speed-up, our approach focuses on relevant degrees of freedom and leverages analytical results.
This method can in principle be extended to more challenging situations in the future.
For instance, droplets embedded in an elastic matrix affect each others growth, which can be described by similar effective theories \cite{VidalHenriquez2021,Vidal2020}.
Similar dynamics will also inform the dynamics of droplets in cells, where for instance chromatin and the cytoskeleton suppress coalescence of condensates \cite{Feric2013,Boeddeker2022,Wiegand2020}.
Our method can in principle also account for fluid flows, which are present in many liquid-like systems~\cite{Brangwynne2009,Setru2021,Brangwynne2011}.
Here, large scale flows will advect the droplets and also affect the background field.
Finally, we could account for Brownian motion of droplets, their coalescence upon contact, and their spontaneous division in sufficiently strongly driven systems~\cite{Zwicker_nature_2016,Seyboldt_2018}.
In particular, droplets have shown anomalous coarsening behavior in cells, primarily due to hindrance and physical barriers which curb their ability of coalesce \cite{Feric2013,Quiroz2020,Boeddeker2022,Lee2021_condensates}.
Extending our effective method to account for these physical processes will allow analyzing more and more complex situations, approaching the complexity necessary to understand the behavior of many droplets in biological cells.

In summary, we have demonstrated that our effective method faithfully captures the effects of chemical reactions and external chemical gradients. %
The method is several orders of magnitude faster than traditional continuous models, making it viable for fast and computationally efficient simulations of systems with many droplets.
More importantly, our model provides a modular platform, which can be extended with other relevant physical phenomena affecting droplets, thus shedding insights on the formation, dissolution, stability, and sizes of biomolecular condensates.

\begin{acknowledgments}
The authors thank Jan Kirschbaum and Swati Sen for helpful discussions.
All authors acknowledge funding from the Max Planck Society.
The authors declare that they have no conflict of interest.
\\
\end{acknowledgments}

\appendix

\section{Interfacial velocity of droplets}

\label{sec:interfacial_velocity}

We investigate the dynamics of the droplet interface by analyzing an infinitesimal cuboid placed on the interface.
The cuboid is aligned with the interface and covers an area element~$\diff A$ of it.
It protrudes by distances $\epsilon_\mathrm{in}$ and $\epsilon_\mathrm{out}$ inside and outside the interface, respectively, so its volume is given by $(\epsilon_\mathrm{in} + \epsilon_\mathrm{out}) \, \diff A$.
Keeping the cuboid fixed in space, the distances change by the interfacial speed $v_\mathrm{n}$ in the normal direction, $\partial_t \epsilon_\mathrm{in} = - \partial_t \epsilon_\mathrm{out} = v_\mathrm{n}$.
We use this to express the time derivative of the amount of material in the cuboid, $\Phi = ( \epsilon_\mathrm{in} \, \phiEqIn + \epsilon_\mathrm{out} \, \phiEqOut )\, \diff A$, as
$    \partial_t {\Phi} = \bigl(
    	\phiEqIn v_\mathrm{n} + \epsilon_\mathrm{in} \partial_t \phiEqIn
	-  \phiEqOut v_\mathrm{n} +  \epsilon_\mathrm{out} \partial_t \phiEqOut
	\bigr) \, \diff A$.
On the other hand, the amount~$\Phi$ changes due to material fluxes, $\partial_t \Phi \approx ( \jIn - \jOut) \cdot \vec{n} \, \diff A$, where we neglect chemical reactions in the infinitesimal volume element and divergences of fluxes tangential to the interface.
Equating the two expressions for $\partial_t \Phi$ in the limit of small $\epsilon_\mathrm{in}$ and $\epsilon_\mathrm{out}$, we arrive at \Eqref{eqn:interfacefluxes}.

\section{Growth and drift of droplets}

\label{sec:droplet_dynamics}

We here derive the expressions for the growth speed and drift rate of droplets of arbitrary shape.
For simplicity, we focus on $d=3$ dimensions, but the derivation works analogously in all dimensions.
The 2D surface of a 3D droplet can be parameterized by points $\vec{R}(\theta, \varphi)$ with parameters $\theta$ and $\varphi$.
For simplicity, we consider convex droplet shapes, so we can place a spherical coordinate system inside the droplet and write $\vec{R}(\theta, \varphi,t) = P(\theta, \varphi, t) \, \vec{e}_r$, where $\theta$ and $\varphi$ denote the typical angles, $\vec{e}_r$ is the radial unit vector of the droplet surface, and $P(\theta, \varphi, t)$ denotes the distance of the surface from the origin.
The droplet volume reads $V = d^{-1} \int P^{\, d} \, \diff\Omega$, where $\diff\Omega$ is the solid angle element, which reads $\diff\Omega = \sin\theta \, \diff\theta\diff\varphi$ for spherical coordinates.
The volume changes in time as $\partial_t V = \int (\partial_t P) \, P^{\, d-1} \, \diff\Omega$, where $\partial_t P$ is the interfacial speed in the radial direction.
We obtain it from the normal speed $v_\mathrm{n}$, given by \Eqref{eqn:interfacefluxes}, using the geometric relation $v_\mathrm{n} = (\partial_t P) \, \vec{e}_r \cdot \vec{n}$, which implies $\partial_t P = v_\mathrm{n} / (\vec{e}_r \cdot \vec{n})$.
Additionally, the differential element~$\diff\Omega$ can be linked to the surface area element~$\diff A$ by $P^{\, d-1}\diff\Omega = \vec{e}_r \cdot \vec{n} \, \diff A$.
Taken together, the rate of change of the radius~$R$ of a sphere with volume~$V$, given by $\partial_t R = (\partial_t V)/S$ for the surface area~$S$, can then be expressed as  \Eqref{eqn:DropletGrowth}.

Similarly, we analyze the center-of-mass position $\vec x = [(d+1)V]^{-1} \, \int P^{\, d+1} \vec{e}_r \, \diff\Omega$.
We find $\partial_t(\vec x V) = \int \! P^{\, d}  \, (\partial_t P) \,  \vec{e}_r  \, \diff\Omega = \int \! P v_\mathrm{n} \vec{e}_r \, \diff A$.
If the droplet is initially centered on the origin, $\vec x =0$, this implies  $\partial_t \vec x = V^{-1} \int \! P v_\mathrm{n} \vec{e}_r\, \diff A$.
This expression is equivalent to \Eqref{eqn:DropletDrift} for spherical droplets, where $P(\theta, \varphi)=R$, $\vec{e}_r=\vec{n}$, and $R/V = d/S$.

\section{Fluxes inside the droplet}

\label{sec:fluxes_inside_droplet}

We determine the flux inside the droplet from the stationary form of the volume fraction field $\phi_\mathrm{in}$ inside the droplet.
We thus solve for the stationary state of \Eqref{eqn:RD_droplet} in a coordinate system with angular symmetry centered at the droplet, so that fields only depend on the radial coordinate~$r$.
Considering the boundary conditions $\phi_\mathrm{in}(R) = \phiEq_\mathrm{in}$ and $\partial_r \phi_\mathrm{in}(0) = 0$, as well as $R \ll \xi_\mathrm{in}=\sqrt{D_\mathrm{in}/|k_{\mathrm{in}}|}$, we find for  1, 2 and 3 dimensions
\begin{subequations}
\label{eqn:fluxes_outside_appendix}
\begin{align}
    \phi_\mathrm{in,1 \mathrm{D}}(r) &= \phi^0_\mathrm{in} + \frac{s(\phi^0_\mathrm{in})}{k_\mathrm{in}} - \frac{s(\phiEqIn)}{k_\mathrm{in}} \frac{\cosh \left(\frac{r}{\xi_\mathrm{in} }\right)}{\text{cosh}\left(\frac{R}{\xi_\mathrm{in} }\right)}%
\\[10pt]
    \phi_\mathrm{in,2 \mathrm{D}}(r) &= \phi^0_\mathrm{in} + \frac{s(\phi^0_\mathrm{in})}{k_\mathrm{in}} - \frac{s(\phiEqIn)}{k_\mathrm{in}} \frac{ I_0\left(\frac{r}{\xi_\mathrm{in} }\right)}{I_0\left(\frac{R}{\xi_\mathrm{in} }\right)} %
\\[10pt]
    \phi_\mathrm{in,3 \mathrm{D}}(r) &= 
    	\phi^0_\mathrm{in} + \frac{s(\phi^0_\mathrm{in})}{k_\mathrm{in}} 
	- \frac{R}{r} \frac{s(\phiEqIn)}{k_\mathrm{in}} \frac{\sinh \left(\frac{r}{\xi_\mathrm{in} }\right)}{ \text{sinh}\left(\frac{R}{\xi_\mathrm{in}}\right)}%
    \;,
\end{align}
\end{subequations}
where $I_0$ is the modified Bessel function of the first kind.
In all dimensions, the fluxes inside the interface are approximately $\vec{j}_\mathrm{in} \approx [-D_\mathrm{in} \boldsymbol{\nabla} \phi_\mathrm{in}(R)] \, \vec{n} \approx \frac{R}{d} \, \left [ s(\phi^{0}_\mathrm{in}) - k_{\mathrm{in}}(\phi^\mathrm{eq}_\mathrm{in} - \phi^{0}_\mathrm{in}) \right ] \, \vec{n}$, where $d$ is the space dimension.

\section{Fluxes in droplet shell sectors}

\label{sec:flux_expressions}

We consider a droplet of radius $R$ with a shell of thickness $\ell$.
To determine the profile $\phiOut^{(m)}(r)$ in the $m$-th shell sector, we solve \Eqref{eqn:RD_dilute} in stationary state.
We assume spherical symmetry and impose the boundary conditions $\phiOut^{(m)}(R)=\phiEqOut$ and $\phiOut^{(m)}(R + \ell)= \phiShell^{(m)}$.
For simplicity, we linearize the reaction flux, $s(\phiOut^{(m)}) \approx \Gamma_\mathrm{out} - \kOut \, \phiOut$, where $\Gamma_\mathrm{out} = [\phiShell^{(m)} \, s(\phiEqOut) - \phiEqOut \, s(\phiShell^{(m)})] / (\phiShell^{(m)} - \phiEqOut)$ and $\kOut = [s(\phiEqOut) - s(\phiShell^{(m)})] / (\phiShell^{(m)} - \phiEqOut)$.
Using the diffusivity\,$D_\mathrm{out}$, we can then determine the local fluxes at the droplet surface in the normal direction, $\jOut^{(m)} \cdot \vec{n} = -D_\mathrm{out} \partial_r \phiOut^{(m)}|_{r=R}$, which in $1$, $2$, and $3$ dimensions read
\begin{widetext}
\begin{subequations}
\begin{align}
    \vec{j}^{(m)}_\mathrm{out, 1D} \cdot \vec{n}
    &= D_\mathrm{out} \frac{
    	(\phiEqOut \, \kOut - \Gamma_\mathrm{out}) \cosh \left(\frac{\ell}{\xi_\mathrm{out}}\right)
	 -\left(\phiShell^{(m)} \, \kOut- \Gamma_\mathrm{out} \right)
}{\kOut \, \xi_\mathrm{out} \, \sinh \left(\frac{\ell}{\xi_\mathrm{out} }\right)}
\\[10pt]
    \vec{j}^{(m)}_\mathrm{out, 2D} \cdot \vec{n}
    &= D_\mathrm{out} \frac{
    (\phiEqOut \, \kOut - \Gamma_\mathrm{out}) \left[I_1\left(\frac{R}{\xi_\mathrm{out} }\right) K_0\left(\frac{\ell+R}{\xi_\mathrm{out} }\right)
    	+K_1\left(\frac{R}{\xi_\mathrm{out} }\right) I_0\left(\frac{\ell+R}{\xi_\mathrm{out}}\right)\right]
	- \frac{\xi_\mathrm{out} }{R} \left(\phiShell^{(m)} \, \kOut - \Gamma_\mathrm{out}\right)
    }
    {\kOut \, \xi_\mathrm{out}  \left[K_0\left(\frac{R}{\xi_\mathrm{out} }\right) I_0\left(\frac{\ell+R}{\xi_\mathrm{out} }\right)-I_0\left(\frac{R}{\xi_\mathrm{out} }\right)
   K_0\left(\frac{\ell + R}{\xi_\mathrm{out} }\right)\right]}
\\[10pt]
    \vec{j}^{(m)}_\mathrm{out, 3D} \cdot \vec{n}
    &=  D_\mathrm{out} \frac{
    	(\phiEqOut \, \kOut - \Gamma_\mathrm{out}) \left(R \coth \left(\frac{\ell}{\xi_\mathrm{out} }\right) + \xi_\mathrm{out} \right)
	- \frac{\ell + R}{\sinh \left(\frac{\ell}{\xi_\mathrm{out} }\right)} \left( \phiShell^{(m)} \, \kOut - \Gamma_\mathrm{out}\right)
	} {\kOut \, \xi_\mathrm{out} \, R},
\end{align}
\end{subequations}
\end{widetext}
where we introduced the reaction-diffusion length scale $\xi_\mathrm{out} = \sqrt{D_\mathrm{out}/|\kOut|}$, and $I_\alpha$ and $K_\alpha$ are the modified Bessel functions of the first and second kind, respectively.
Note that the definitions for $\Gamma_\mathrm{out}, \, \kOut$ diverge when $\phiShell^{(m)} \approx \phiEqOut$.
We then simply approximate the reaction flux, $s(\phiOut^{(m)}) \approx \overline{\Gamma}_\mathrm{out}$, where $\overline{\Gamma}_\mathrm{out} = [s(\phiShell^{(m)}) + s(\phiEqOut)]/2$.
As before, we solve \Eqref{eqn:RD_dilute} in stationary state assuming spherical symmetry along with the boundary conditions $\phiOut^{(m)}(R) = \phiEqOut$ and $\phiOut^{(m)}(R + \ell) = \phiShell^{(m)}$.
After obtaining $\phiOut^{(m)}$ inside the shell, the fluxes $\jOut^{(m)} \cdot \vec{n}$ in 1, 2 and 3 dimensions then read as
\begin{widetext}
\begin{subequations}
\begin{align}
    \vec{j}^{(m)}_\mathrm{out, 1D} \cdot \vec{n}
    &= \frac{D_\mathrm{out} \,  (\phiEqOut - \phiShell^{(m)})}{\ell}-\frac{\overline{\Gamma}_\mathrm{out} \, \ell}{2}
\\[10pt]
    \vec{j}^{(m)}_\mathrm{out, 2D} \cdot \vec{n}
    &= \frac{\overline{\Gamma}_\mathrm{out} \, \ell \, (\ell+2 R) - 4 \, \phiEqOut \,  D_\mathrm{out} + 4 \, \phiShell^{(m)} \, D_\mathrm{out}}{4 \, R \, \log \left(\frac{R}{\ell+R}\right)}+\frac{\overline{\Gamma}_\mathrm{out} \, R}{2}
\\[10pt]
    \vec{j}^{(m)}_\mathrm{out, 3D} \cdot \vec{n}
    &=  -\frac{\overline{\Gamma}_\mathrm{out} \, \ell^2 \, (\ell+3 R) - 6 \, \phiEqOut \, D_\mathrm{out} (\ell+R) + 6 \, \phiShell^{(m)} \, D_\mathrm{out} (\ell+R)}{6 \, \ell \, R}.
\end{align}
\end{subequations}
\end{widetext}

\bibliography{references.bib}

\end{document}